\documentstyle[preprint,aps,prc,12pt,epsfig,rotating,subfigure]{revtex}
\begin{document}
\draft
\title{ FUSION AND BINARY-DECAY MECHANISMS 
 IN THE $^{35}$Cl+$^{24}$Mg SYSTEM AT E/A $\approx$ 8
 MeV/NUCLEON} 

\author{ Sl. Cavallaro$^{1,2}$, E. De Filippo$^{o}$, G. Lanzan\`o$^{o}$
, A. Pagano$^{o}$ 
and M.L. Sperduto$^{1,2}$}
\address{ $^{1}$Universit\`a di Catania and $^{o}$INFN, Corso Italia, 57 95129 
Catania}
\address{  Laboratorio Nazionale del Sud$^{2}$, Via S. Sofia, 44 95123 Catania,
,Italy}
 \author{ R. Dayras, R. Legrain and E. Pollacco}
\address{CEA,DAPNIA/SPhN CE-Saclay, F-91191 Gif-sur-Yvette CEDEX, 
France}
\author{ C. Beck, B. Djerroud$^{*}$, R.M. Freeman, F. Haas, 
A. Hachem$^{+}$, B. Heusch,  D. Mahboub,  A.Morsad$^{\#}$ and R. Nouicer}
\address{Institut de Recherches Subatomiques, {\it IN2P3/Universit\'e
Louis Pasteur}, F-67037 Strasbourg, France}
\author{ S.J. Sanders}
\address{ Department of Physics and Astronomy, The University of 
Kansas, Lawrence KS 66045}

\address{ * Department of Chemistry, University of Rochester,Rochester, 
NY 14627, USA}
\address { + Permanent address: Facult\'e des Sciences, Universit\'e 
de Nice, F-06034 Nice, France}
\address {\# Permanent address: Facult\'e des Sciences, Universit\'e
Hassan II, Casablanca, Maroc}

\date{October, 1997}
\maketitle

\begin{abstract}
Compound-nucleus fusion and binary-reaction mechanisms have been
investigated for the $^{35}$Cl+$^{24}$Mg system at an incident beam 
energy of
E$_{Lab}$= 282 MeV. Charge distributions,  inclusive energy spectra,
and angular distributions have been obtained for the evaporation 
residues and the binary fragments.
 Angle-integrated cross sections have been determined for evaporation 
residues from
both  the complete and incomplete fusion mechanisms. Energy spectra for
binary fragment channels near to the entrance-channel mass partition are 
characterized by an inelastic contribution that is in addition to a fully
energy damped component.
The fully damped
component which is observed in all the binary mass channels can be associated 
with decay times that are comparable 
to, or  longer
than the rotation period. 
 The observed mass-dependent cross sections for the fully
damped component are well reproduced by the fission transition-state
model, suggesting a fusion followed by fission origin.  The
present data cannot, however, rule out the possibility that
a long-lived orbiting mechanism accounts for part or all of this yield.


\end{abstract}

\pacs{ 25.70.-z}


\newpage

I. INTRODUCTION

  In recent years heavy-ion induced reactions, involving
intermediate mass systems as light
 as A$_{CN}$$\leq$60 at bombarding energies  $\leq10$ MeV/nucleon, 
  have been studied for  various target+projectile combinations over
 a wide  energy range [1-8] . Near the Coulomb barrier (region 
 I), the complete fusion process (CF)  is the dominant reaction 
mechanism. At  higher energies (region II) however, this process  is
 limited by the contributions of other, competing processes such as quasi- 
and deep-inelastic collisions (DIC).
 The CF cross section is  still  increasing with energy,  but at a much slower
rate than in region I.  At even higher energies (region III) the general 
instability of the composite system leads to a decreasing CF cross section with
increasing beam energy. Strong competition  between fusion and other damped
mechanisms characterize regions II and III. 
   The present measurement, which has been performed at an incident energy
corresponding to region III for the complete-fusion cross sections, is designed
to further study the competing reaction mechanisms at higher energies. 
Several competing mechanisms have been suggested to absorb the reaction flux in
regions II and  III. Heavy-ion resonances and orbiting mechanisms have been
shown to compete with fusion for the incident flux in region  II [9],at least 
in some systems. 

In the incomplete fusion process (ICF) only part of the entrance channel mass 
is incorporated into the resulting compound system. While this process is
insignificant at lower energies [10,11], it can quickly rise at higher energies
in region III, leading to a large reduction of the complete fusion cross
section [10].

Classical trajectory models that include consideration of frictional forces and
the effects of  thermal fluctuations, can be used to predict observables such
as the mean values and widths of the mass, charge and energy flow in damped,
deep-inelastic collisions [12,13]. By the Langevin Method (LM), and on the
basis of the surface friction model of ref.[14], for instance, it is further
possible to model the competition between fusion and DIC mechanisms by
introducing temperature-dependent coefficients for the radial, tangential and
deformation friction contributions, respectively. However, the determination of
these coefficients to the precision needed to reproduce the fluctuations around
 dynamical central values has only been accomplished for a few cases. More
experiments are needed where the excitation functions for fusion and DIC are
measured simultaneously for a given nuclear system. 

The present measurement  is part of a more general experimental  program 
devoted to the study of the limits of compound  nucleus formation by measuring 
the macroscopic observables related to the formation and decay of a single
nuclear system through several different entrance channels.

 Within the LM method a limit is reached when  a given trajectory starts to be
captured in the pocket of the ion-ion potential. At this point there is no
further information that can be determined for the dynamical trajectories
[15,16], and the two interacting nuclei can be thought to form a di-nuclear
system (DNC), from which they can either fuse into a fully equilibrated
compound nucleus (CN), or escape from the ion-ion potential well, producing a
damped orbiting process. 

The decay properties of the binary fragments produced as a consequence of the
fission after fusion [17] or after scission of a long-lived orbiting di-nuclear
system [18] have been described in the literature on the basis of the
phase-space configurations at the saddle point of the intermediate system. It
has been found to be difficult to experimentally distinguish between the
fission and orbiting mechanisms as both are expected to give very similar
behaviour [19]. A fusion-fission origin has been suggested for fully-damped
yields associated with reactions populating the  $^{56}$Ni [1],  $^{58}$Ni [2],
and $^{47}$V [3] compound nuclei.  On the other hand similar binary yields
observed for certain lighter systems have been alternatively explained  in
terms of a statistical di-nucleus orbiting model [4-6].

This paper reports on the results of both the fusion-evaporation  and binary
fragment  ( fusion-fission and DIC ) cross sections  for the 
$^{35}$Cl+$^{24}$Mg reaction at E$_{Lab}$= 282 MeV, obtained by inclusive
energy and angular distribution measurements as functions of the nuclear charge
of the detected fragments. Previously, this reaction has been studied only at
lower energies [20] and only few results have been published in a rapid
communication, [21]. However, the composite system reached by this reaction, 
$^{59}$Cu ,  has been extensively investigated in fusion regions II and III  by
the $^{19}$F+$^{40}$Ca [22] and $^{32}$S+$^{27}$Al  reactions [22-27]. These
previous measurements have shown  pronounced entrance-channel effects for the
relative contributions of DIC and the ICF process. However fission yield
measurements were lacking.  It is then also interesting to study the formation
of the same $^{59}$Cu nucleus in conditions of an intermediate entrance-channel
mass asymmetry.

Complementary information concerning the mean excitation energies and average
angular momenta of  binary-fragments from the $^{59}$Cu intermediate system
formed  by the $^{35}Cl+^{24}Mg$ reaction  will be reported in a forthcoming
paper based on in- and out-of-plane angular correlations between light
particles and heavy fragments [28]. 

 In the next section, the experimental procedure is described.   In section
III, the fusion-evaporation yields are given and discussed. The properties of
the binary fragments  are presented in section IV.  In Section V, the 
compound-nucleus-decay models used to discuss the data are briefly introduced,
and comparisons of the binary fragment experimental results with the
predictions from the fission transition-state model are presented. In section
VI, the present data are discussed in the context of previously reported
results and used to explore how different competing mechanisms lead to
limitations in the formation of the $^{59}$Cu compound nucleus. Finally, a
summary of the  main results and  conclusions are given in Section VII.  A
recent paper  has presented the preliminary  results obtained from the 
fragment-fragment angular correlation measurements performed at E$_{Lab}$=278.4
MeV [29], which confirm well the main conclusions of this works and provides
some information on the possible occurrence of three-body break-up processes. 

\newpage

 II. EXPERIMENTAL PROCEDURE

The experiment was performed by  Saclay Post-Accelerator Tandem Facility. A
$^{35}$Cl pulsed beam ( charge state Z=14$^{+}$ )  accelerated to an energy
E$_{Lab}$=282.4 MeV  was focussed onto a self-supporting rolled $^{24}$Mg
target, located in a 2 meter diameter scattering chamber. The  $^{24}$Mg foils,
of 99.9\% isotopic enrichment,  had areal densities of  350$\pm$32 and
255$\pm$20 $\mu$g/cm$^{2}$, respectively. The time structure of the pulsed beam
had a period of  37 ns.  The beam current was varied from a few nA to
$\approx$20 nA, depending on the positions of the detectors and the
corresponding counting  rates. Carbon and oxygen contaminants were
experimentally estimated at $\leq$10$\mu$g/cm$^{2}$ each, using a 2 MeV
$\alpha$-beam backscattering-technique. Heavy fragments ( Z=13-26 ) were
detected by a Bragg-curve ionization chamber (BIC)  filled with CF4 gas at a
pressure of 150 torr in the  3$^{o}$ to 12$^{o}$ angular range with a  1$^{o}$
step increment. Four small gas-ionization chambers, at a pressure of 51 torr of
CF4  gas, followed by silicon E-detectors, were used to detect fragments with
Z=3 to Z=26 in the  10$^{o}$-90$^{o}$ angular range with a 2$^{o}$ step.  An
additional gas-silicon telescope was located at a fixed position,  10$^{o}$
with respect to the beam as a monitor.  The beam current was  measured  in a
Faraday cup and integrated. To correct for carbon contamination, measurements
were also made at the same detector angles and with comparable beam conditions
using a 100$\mu$g/cm$^{2}$ thick, self-supporting,  carbon foil as a target
[30]. Absolute cross sections were determined by measuring the elastic
scattering of 282.4 $^{35}$Cl  ions from the $^{24}$Mg target at
$\theta_{Lab}$=3$^{o}$-16$^{o}$ and comparing with the elastic scattering
predictions. The optical model analysis of the experimental data performed with
the code PTOLEMY is shown in Fig.1.  This calculation was made using the
following parameters, V$_{R}$=5.1 MeV, r$_{R}$=1.46 fm, a$_{R}$=.5 fm,
W$_{I}$=9.3 MeV, r$_{I}$=1.3 fm, a$_{I}$=.42 fm  for the real and imaginary
parts of the potential respectively  and a coulomb radius parameter r$_{C}$=1.2
fm. These parameter values have been extrapolated from the low-energy data of
ref. [31]. 

Taking into account the uncertainties associated with all of the measured
parameters, a global error of 17\% is attributed to the absolute values of the
differential cross sections reported in this paper. Double differential cross
sections determined from the experiment have been corrected for the
contribution of the carbon contaminant by  subtracting from the measured cross
sections the component : 

\begin{equation}
{(\frac{d^{2}\sigma}{{d\Theta}dE})}_{corr}={(\frac{d^{2}\sigma}
{{d\Theta}dE})}_{C}\frac{t_{cont}}{t_{carb}}
\end{equation}

where  (d$^{2}\sigma$/d$\Theta$dE)$_{C}$ is the  cross section determined for
the reaction $^{35}$Cl+$^{12}$C  [30], t$_{carb}$ is the carbon-target
thickness  and t$_{cont}$,  the carbon contamination on the $^{24}$Mg target. 
These corrections are generally $\leq$10\%.
However, because of their very different kinematics, reactions induced on the
carbon contaminant can contribute significantly to the yield of some fragments
at particular angles and energy ranges. The energy calibrations of the BIC
detector and of the gas-silicon telescopes were determined by measuring the
peak positions of the elastically scattered  $^{35}$Cl nuclei from a 
400$\mu$g/cm$^{2}$ thick, gold target  and using the known kinematics for this
reaction. In the energy calibration procedure, corrections for energy losses in
the $^{24}$Mg target, ( in the half-thickness approximation and with exception
of heaviest residues, see later),  and for energy losses of the fragments
passing through the mylar windows of the BIC and the IC were included.
Pulse-height-defect corrections for the Si detectors were taken into account
using the procedure given in [32]. 

\bigskip
\bigskip
 
III. FUSION-EVAPORATION CROSS SECTIONS

The dominant mechanism by which the compound system de-excites in this
reaction, at least for lower spin values,  is through the emission of light
particles.  Cross sections for the  evaporation-residues arising from both the
complete fusion and possibly by incomplete fusion processes were therefore
determined. Energy spectra obtained at  $\theta$$_{Lab}$=7$^{o}$ are shown in
Fig. 2. The bell-shaped patterns characteristic of evaporation residues can be
seen for Z=24 to Z=21  fragments. These data are compared to  statistical-model
 calculations with the  code LILITA [33] 

In making the comparison with model calculations it  is important to account
for fragment energy losses in the target. For the heavier fragments these
losses can amount to 7 MeV, as compared to a $^{35}$Cl incident beam energy 
loss of only 2.6 MeV. The calculated spectra in Fig. 2 were obtained by
considering the reaction  as occurring  in  a stack of twenty  thin-``targets"
with a total thickness equivalent to the actual $^{24}$Mg target used in the
experiment. The LILITA spectra were calculated for each of the twenty
``targets"  and then corrected for the energy loss through the remaining
targets. The final, ``degraded" LILITA spectrum was then obtained  by  a
convolution of the twenty constituent spectra. This procedure allows one to
compare the experimental Z-residue spectra with those expected from the full
momentum transfer kinematics. 

For complete fusion events,  the light particles are emitted in the rest frame
of the recoiling nucleus, with angular distributions symmetric around 90$^{o}$
and with Maxwellian energy spectra. If, in addition,  the angular distributions
are assumed to be isotropic, then the energy spectra of the evaporating
residues can  be expressed by [33] : 

\begin{equation}
{(\frac{d^{2}\sigma}{d{\Omega}{dE_{ER}}})=\frac{K{(2E_{ER})^{1/2}}}
{M^{3/2}_{ER}}{\times}e^{-\frac{E_{CN}sin^{2}\theta_{Lab}}{s^2M_{CN}}
}{\times{e^{-\frac{(E^{1/2}_{ER}-E^{1/2}_{C})^2}{s^2M_{CN}}}}}}
\end{equation}

where E$_{ER}$ and M$_{ER}$ are the kinetic energy and the mass of the
evaporating fragment, respectively, E$_{CN}$ is the kinetic energy of the
compound system, K is a normalization factor and $s$ is the standard deviation
parameter of the Gaussian velocity distribution of the recoiling residues. For 
M$_{ER}$, the mean values of isotope mass-distributions calculated by LILITA,
have been assumed. 

The E$_{C}$ term is given by:
\begin{equation}
E_{C}={\frac{M_{ER}E_{CN}}{M_{CN}}{\cdot}cos^{2}\theta_{Lab}}
\end{equation}

If the E$_{ER}$ energies  are large compared with the ${s^2M_{CN}}$ term, then 
the energy variation resulting from the E$_{ER}^{1/2}$ term can be neglected
compared to that arising from the  exponential term and the resulting energy
spectra predicted by Eq.(2)  are centered at the energies given by Eq.(3).  The
experimental observation of  energy spectra centered at these energies is then
a signature of full momentum transfer. Eqs.  (2) and (3) are no longer exact if
the  angular distributions of  evaporated particles are anisotropic,  however
departures from Eq.(2) are expected to be relatively small, especially in
reverse kinematics conditions. In the present case, the Eq. (3) values are
found to be lower than  residue energy spectrum centroids calculated by the
LILITA code with anisotropic angular distributions by only 1 to 1.5 \%.  Then,
for the present reaction, the centroid energies obtained using  Eq. (3) are a
good measure of the extent to which incomplete momentum-transfer processes
occur. 

In Fig. 2, the predicted centroids from Eq. (3) are shown by arrows.  The
excellent  agreement with the observed centroids  indicates that there is no
appreciable contribution of incomplete fusion processes for fragments with
22$\le$Z$\le$24. For instance, if a pre-equilibrium  $\alpha$ particle was
``lost" in the reaction by either the projectile or  target before the fusion
process occurs, centroid energy shifts  of $\approx$-14 MeV or $\approx$10 MeV,
respectively,  would be expected. Since  no shift was found, these fragment
yields have been ascribed to  the complete fusion mechanism. It should also  be
noted that the LILITA calculations reproduce quite well both the energy 
centroid and the shape of the experimental spectra for these   fragments. This 
justifies using these calculations to discriminate between the complete fusion
process and other reaction mechanisms.

A departure from the bell-shaped behaviour is clearly seen for the fragments
with Z=20 and, to a lesser extent, also for Z=21. The deviation from the
compound-nucleus evaporation behavior becomes  very large for fragments with
Z$\le$19, indicating that  faster mechanisms  other than complete fusion are
contributing to these yields [see below]: quasi-, deep-inelastic processes. 

Evaporation-residue cross sections of complete-fusion were  generally
determined for fragments with Z$>$21,  by integration of experimental angular
spectra. For $\theta_{Lab}$$<$3$^{o}$, and  only in few cases where
identification thresholds prevented to use the experimental data ( as at most
forward angles for Z=24 ), LILITA calculations were used to extrapolate the
angular distribution after normalization to the experimental data at a larger
angle. 

 To better fit the energy spectra of 18$\le$Z$\le$21 residues, at least at the
most forward angles, it was necessary to introduce a contribution from an
incomplete fusion process characterized by a loss of mass of $\approx$8 amu
from the target. This effect was simulated by calculating  evaporation-residue
spectra from the  reaction $^{35}$Cl+ $^{16}$O using the LILITA code and
assuming a critical  angular momentum of 28$\hbar$. For those spectra where a
centroid shift was evident,  the heights of both the complete- and
incomplete-fusion components were adjusted simultaneously in order to obtain
the best overall reproduction of the data. The decomposition in the CF and ICF
contributions is shown in Fig.2 and also in Fig.3  for Z=20 residues detected
at four different angles. For Z=19 and Z=18 residues the presence of deep
inelastic components increased  the difficulty of separating the various
reaction components. 

Experimental angular distributions including all mechanism contributions are
shown in Fig.4 for fragments with 16$\le$ Z$\le$24. For Z=22 to Z=24 residues,
total yields are attributed to the CF process and compared with LILITA
calculations. For residues with Z=18 to Z=21, contributions from the CF and ICF
processes  have been separated. For the Z=18 and 19 fragments, are reported
both the  experimental total yields, and the 'extracted' CF and ICF
contributions. For the projectile-like Z=16 and 17 fragments, the spectra
become dominated by quasi-elastic and deep-inelastic scattering contributions. 

Evaporation-residue cross sections  from complete-fusion were determined by
integration of associated angular  distributions.  In few cases where
identification thresholds at the most forwards angles prevented to use the
experimental data, LILITA calculations, after normalization to the experimental
data at a larger angles, were used to extrapolate the angular distributions
down to 0$^{\circ}$. Evaporation residues from incomplete fusion were
determined in the same manner. The complete fusion-evaporation cross-section
amounts to 600$\pm$105mb whereas the incomplete fusion-evaporation amounts to
60$\pm$30mb. Thus, the total evaporation residue cross section is found to be
660$\pm$110mb. 
                                                                               
The complete-fusion evaporation-residue cross sections, obtained by integrating
the angular distributions for the different charge channels, after subtraction
of the estimated ICF contribution, are shown in Fig. 5 by the full circles.
Compared to these experimental results are cross sections calculated using the
LILITA  and CASCADE [34] codes $^{a}$. For these calculations, the diffuse 
cut-off
approximation was assumed for the entrance-channel transmission coefficients
[34], taking the  critical angular momentum for fusion as l$_{cr}$= 38$\hbar$
and a diffuseness of $\Delta$ = 1$\hbar$, extracted from  the total measured 
fusion-evaporation cross section. Known low-lying states were used to determine
the fragment level densities at low excitation energy. As seen in the figure, 
both calculations show  an upward shift in Z-values of $\Delta$Z$\approx$2 with
respect to experimental values. 

It has been suggested that this shift may be due, to some extent,  to  the
effects of deformation in the evaporating system, (see i.e.  refs. [23,35]). 

  Taking deformation into account, however, represents a difficult
task.Different procedures have been debated in the literature  in connection
with fitting  of proton and alpha spectra, [36,37,38,39,40,41]. Due to the lack
of  experimental transmission coefficients for emission of light particles from
 hot deformed nuclei and limitations of  available  multi-step evaporating-code
for  deformed nuclei,  this problem is still open and the object of
investigations [42,43]. 
  
 A simple, although schematic, way  of taking into account
 deformation effects is to increase  the potential radius in calculating the 
transmission coefficients for deformed nuclei and also to increase the moment 
of inertia radius parameters used in determining the available phase space 
for light fragment decay. This procedure has been  used in the present 
analysis in order  to take into account of deformation effects;
 as a matter of fact the deformation could, in our case, simulate a similar 
effect as the ICF mechanism  in the residue charge distribution. 


 Data reported in literature  for this mass region and lower excitation 
energy [36], indicate the need to increase the potential radius, $r$,
 by  10\% to 25\% and to introduce an angular momentum dependent Yrast line 
in the fitting procedure  of evaporation spectra, with a moment of inertia 
given by I=I$_{o}$(1+def$\cdot$J$^{2}$+defs$\cdot$J$^{4}$). 
 The ranges for the related parameters are about r$_{oLDM}$=1.15 to 1.28, 
def=1 to 3.2$\cdot$10$^{-4}$ and defs=0 to 20$\cdot$10$^{-8}$, respectively.
Based on the results obtained at higher excitation energy for the same 
$^{59}$Cu nucleus by fragment-light  particle coincidence measurements [28], 
r=r$_{o}$$\cdot$1.10 and  r$_{oLDM}$=1.28 and def=3.2$\cdot$10$^{-4}$ and  
defs=16$\cdot$10$^{-8}$ have been here assumed in the Cascade  calculations 
and r=r$_{o}$$\cdot$1.10 and r$_{oLDM}$=1.28 in the Lilita calculations. 
The resulting charge  distributions are shown in Fig. 5, indicating a better 
agreement between data and theoretical calculations.

 Fig. 6 reports  the evaporation-residue Z-distributions for
 the ICF  contributions together with corresponding LILITA
calculations. When performing these
statistical model calculations, a total ICF cross section of $\sigma$=60 mb  
was assumed.

\bigskip
\bigskip
\bigskip

 IV. BINARY FRAGMENTS

For fragments such that (3$\le$Z$\le$17), the energy spectra
are no longer consistent with the behaviour expected for heavy residues 
arising from a fusion-evaporation process.
 These nuclei are mainly the remnants of quasi-projectile or
quasi-target fragments from binary reactions. They may also result in part
from a fusion-fission process as we will show later. In this case, the
corresponding cross-section should be added to the fusion cross section.
                                                                             
Fig. 7 shows  the double differential cross section  
($d^2\sigma$/d$\Theta_{cm}$d$E_{cm}$) contours, as a function 
of total kinetic energy TKE and angle $\Theta$$_{cm}$ in the 
 center-of-mass system,  for 5$\le$Z$\le$16 fragments.
The data  exhibit a large range of kinetic energies  and angular 
distributions which are strongly depending on both energy and  fragment 
Z-value.
The angular distributions show  generally an initial  
($d^2\sigma$/d$\Theta_{cm}$d$E_{cm}$)-constancy  at low energy, which  is 
progressively lost at higher energies. 
In order to extract more detailed information, each map has been
divided in 5 MeV-wide energy-slices  and separate angular distributions 
generated for each slice.  In some cases a single component dominates the 
entire kinetic energy range. This is seen in Fig. 8 for the light fragments 
with Z=3 to 5. In this figure  the contributions of all energy slices have 
been summed.
One finds for the lightest fragments that the kinetic energies are fully damped 
and the angular distributions are almost flat.  For fragments with Z=13 to 16, 
a forward-peaked and partially-damped component (projectile  like component) 
is present together with a second, angle independent component.
The former becomes very small at energies higher than $\approx$29 MeV. 
 For  fragments with Z=6 to 12, as shown in Fig. 9, a  component with
a constant cross section can be seen up to 25-30 MeV (left side of figure),
 whereas for higher energies, anisotropic components are also present. 
For Z=6 to Z=10, the anisotropic components are backward peaked, 
(target-like component).
For fragments with Z=11 and Z=12, both projectile-like and target-like
components are present, indicating a net transfer of nucleons from both target
to projectile and viceversa.

 The energy-dependence of the angular distributions clearly indicate 
different reaction times, not only for different fragment pairs,
but also for same pair at different net energy loss.
 The decay time estimates for a surface reaction can be deduced 
from the angular distribution analysis using a simple Regge-Pole Model 
described in Ref.44.  This analysis leads however, to almost flat 
distributions for decay times comparable to the rotation time of 
the di-nuclear system.
In cases where there are both long- and short-lived reaction components,
the Regge Model can be modified by the addition of a constant term h to account
for the long-lived component:

\begin{equation}
\frac{d\sigma}{d\Theta_{cm}}=k[e^{-\frac{\Theta_{cm}}{\omega\tau}}+
e^{-\frac{(2\pi-\Theta_{cm})}{\omega\tau}}] + h,
\end{equation}

Here, $\omega$ is the rotational frequency and $\tau$ is the decay time of the
di-nuclear system. The rotational energy and the  angular momentum at the
scission configuration which are related to  $\omega$, can be estimated
following the procedure discussed in Ref.45. 

  The rotation time of the di-nuclear system, T=2$\pi$/$\omega$,  was found to
be $\approx$1.6$\cdot$10$^{-21}$ s.  Fits of angular distributions by equation
(4) are reported in table I. The   $\tau$$_{1}$ parameter refers to the decay
time of the forward-peaked component. 
 To obtain the $\tau$$_{2}$ decay time of the backward-peaked component, 
( target-like behaviour ), the $\Theta$-abscissa has been changed in
$\Theta_2$=$\pi$-$\Theta$. In the fitting procedure,   the fully damped part of
spectra of all fragments was successfully fitted with only a component with
$d\sigma$/d$\Theta_{cm}$=constant, 
 whereas the partially damped components, with TKE$\ge$29 MeV and fragments
with Z=6 to 17 were  
fitted with eq. 4. In these cases, the cross sections corresponding to the 
constant term in eq.4 accounted for only a few percent of the of total yields.

Table I, reports the angle-integrated cross section and kinetic energy 
values
for both fully-damped components and partially-damped components  for fragments
with Z=3 to Z=16.

 Fig. 10, presents the total measured yields for fragments with Z=3 to
Z=25,
the dashed-line histograms correspond to the fully damped  cross sections
for Z=3 to Z=17 fragments obtained by extrapolating the experimental
distribution over the complete angular range.
For the sake of completeness, the figure also shows the contribution from
the complete-fusion process (full line histogram). The excess
yield around Z=16-17 can be imputed to quasi-elastic and partially damped
processes.

\bigskip
\bigskip

V. TRANSITION-STATE CALCULATIONS

Although neutron, proton, and alpha-particle emissions are the dominant 
particle-decay modes for compound nuclei  as formed in the fusion reactions of
low-mass systems,   binary  fission is also a possible process [17].  For
lighter systems that are below  the Businaro-Gallone limit,  the fission
barriers are such as to favor the  breakup of the compound nucleus into
asymmetric-mass channels.  For these systems,  light-particle emissions can be
thought of as  one extreme of the fission mass distribution [46]. 
Fission to the more symmetric-mass binary channels tends to be restricted in
these systems to partial waves near to the critical angular  momentum for
fusion   l$_{cr}$. Calculations using the transition-state model, where the
fission decay width is determined by the density of states  at the saddle
point, have been found to successfully reproduce the observed fission cross
sections in system of mass A$\approx$100 [47] and lighter [17], when the
saddle-point energy is calculated for different mass asymmetries and spins
using the finite-range model (see, e.g., ref. [48]) .   Binary fragment
production is also predicted by the Extended Hauser-Feshbach Method (EHFM)
[49], where the partial decay widths are determined by the available phase
space at the scission, rather than the saddle, configuration.  This method  has
been successfully used to explain the yields of both evaporation residues and
intermediate mass fragments, as well as the  kinetic energies of the heavier
fragments. 

 Since the saddle-point shape in this mass region corresponds to
having two touching spheroids in an elongated configuration,
any process that leads to the formation of a
long-lived, dinuclear orbiting system of similar deformation is likely to 
result in angular and excitation energy distributions that are indistinguishable
from those of a fusion-fission mechanism. The most unambiguous way to  
distinguish between
the fusion-fission and long-lived orbiting mechanisms in light systems is
to establish an entrance channel dependence of the
fully energy-damped yields that cannot be accounted for by differences
in the expected compound-nucleus spin distribution.
Such studies have not been done for the present system.
Based on the number of open channels systematics of Beck et al.[9], however,
a significant orbiting yield is not expected for the
present reaction and  consequently, in the following we will
assume a fusion-fission origin for these yields.

In the present analysis,  the formulation of Ref. [17], with its
basic-parameter values, is used  to calculate the mass-dependent yields and
kinetic energies for the fission fragments. This calculation uses  the spin and
mass-asymmetry dependent saddle-point energies of the finite-range model [48]
to determine the transition-point phase space. 

In light systems the saddle- and scission-point configurations are believed
to be similar. Therefore, the geometry of the two fragments at scission is
based on the calculated saddle point configuration of the finite-range model.

The total kinetic energy in the exit-channel is assumed to be :
\[
   T_{KE}= V_{C} + V_{N} +\frac{\hbar^{2}L_{out}(L_{out}+1)}
{2\Im_{rel}}      \]
 with \[
L_{out}=\frac{\Im_{rel}}{\Im_{tot}}J_{tot}
\]
and
 \[
\Im_{tot}=\Im_{rel} + \Im_{3} + \Im_{4}.
\]
where V${_C}$ and V${_N}$ are the Coulomb and nuclear energies, 
respectively.
   $\Im_{rel}$ is the relative moment of inertia and $\Im_{3}$ 
and $\Im_{4}$
are the moment of inertia of fragments 3 and 4 respectively.

The diffuse cut-off model [34]  was used to determine the fusion partial cross
section distribution, with the diffuseness  and critical angular momentum for
fusion taken as  $\Delta$=1$\hbar$ \ \ and l$_{cr}$=38$\hbar$, respectively, as
assumed for the evaporation residue analysis presented earlier.   The ratio of
the fission decay width $\Gamma_{f}$ to the total  decay width of the compound
nucleus, $\Gamma_{tot}$, was determined using the statistical model,  with 

 $\Gamma$$_{tot}$ =$\Gamma$$_{n}$ +$\Gamma$$_{\alpha}$ +$\Gamma$$_{p}$
 +$\Gamma$$_{f}$ ,
and
\begin{equation}
\Gamma_{f}=\sum^{A_{CN}/2}_{A_{L}=6}\sum^{A_{CN}/2+2}_{Z_{L}=A_{L}/2-2}
\Gamma_{f}(Z_{L},A_{L})  .
\end{equation}

where $\Gamma$$_{f}$(A$_{L}$,Z$_{L}$) is the decay width for the  channel
specified by the lighter fragment of charge Z$_{L}$ and mass A$_{L}$.   The 
branching ratios and fission cross sections are calculated using
energy-integrated widths for compound-nucleus decay. ( See ref.17 for a more
complete discussion).

Since it is possible that fission events will leave the resulting fragments at
excitation energies above their particle-emission thresholds, secondary
light-particle emission  can affect the mass and energy distributions of the
observed fragments. In order to make a realistic comparison  of the
transition-state model calculations to the experimental  results, this
secondary light-particle emission has been simulated   by using the binary
decay option of the LILITA code [33]. The cross sections for the primary mass
distribution, obtained from the transition-state model  calculations, have been
taken as input data. In each fission exit channel, a Gaussian distribution was
assumed for the total kinetic energy distribution, with the peak of the
distribution obtained using the double spheroid model and the standard
deviation taken as 21\% of the average value.  The total excitation energy was
divided between the two fragments assuming equal temperatures.

 Comparisons of transition-state model predictions with the experimental 
cross sections 
are  shown in Fig. 11 for all Z-fragment yields in the range 3$\leq$ Z$\leq$25.
For 3$\leq$ Z$\leq$17 only the cross sections based on the fully damped yields 
are shown.
For 18$\leq$ Z$\leq$25 the CF experimental yields  are reported. 
It is worth noting  the  large shift to lower values that the 
evaporation process produces in the final fragment-distribution with respect to
the primary one. Good overall agreement is found between the experimental 
results and the transition-state  calculations.  
The calculated fission cross section is 145 mb whereas 
the corresponding isotropic experimental cross section of 168$\pm$30mb.

The calculated and experimental c.m. total  kinetic energies  
of the fragments are  compared in Fig. 12.
The transition-state model calculations corrected for evaporation are  
generally in good agreement with the measured kinetic energies.

Then the overall agreement suggests to ascribe the fully-damped yields to 
the fusion-fission process.

\bigskip
\bigskip

VI. LIMITS TO THE $^{59}$Cu COMPOUND NUCLEUS FORMATION

The energy dependence of the total fusion cross section can yield important
information about the limits of energy and spin that can be sustained by
nuclear systems. Fusion cross sections for the  $^{35}$Cl+$^{24}$Mg (present
work and Ref. [20]),  $^{19}$F+$^{40}$Ca [22] and  $^{32}$S+$^{27}$Al [22-27]
reactions are shown in Fig. 13 as functions of V$_C$/E$_{c.m.}$. The Coulomb
barrier energy V$_C$ is taken as 

\begin{equation}
{V}_{C}\ =\ 1.44\ {{Z}_{1}{Z}_{2} \over {R}_{B}}\left({1\ -\ {0.63 
\over{R}_{B}}}\right)
\end{equation}

with
$${R}_{B}\ =\ 1.07\ \ \left({{A}_{p}^{1/3}\ +{A}_{t}^{1/3}}\right)\
 +\ 2.73 (fm)$$
A$_p$ and A$_t$ are the projectile and target mass respectively, in atomic mass
units, and R$_B$ is the barrier radius. Each of these reactions populate the
$^{59}$Cu compound nucleus. The energy dependence of the total reaction cross
section,  based on optical model calculations for the $^{35}$Cl+$^{24}$Mg
system (see caption of Figure 1 ),  is shown  by the long-dashed line in Fig.
13.  At energies near and somewhat  above the barrier,
V$_C$/E$_{c.m.}$$\geq$0.6  (region I), the fusion cross section is close to the
total reaction cross section, except for small departures that result from
quasi elastic processes.   In the range 0.3$\le$V$_C$/E$_{c.m.}$$\le$0.6
(region II) the fusion cross section initially continues to increase with
incident projectile energies, though at a lower rate,  and then reaches a
saturation point. Competing with fusion, in this energy range, are
deep-inelastic processes that are strongly dependent on the entrance-channel
mass asymmetry. Region II can be described in terms of the critical distance
model [49]  or by dynamical trajectories based on friction models. Estimations
of  Bass model [50] for disappearance of pocket in the ion-ion potential 
( l$_{fus}$$\approx$38$\hbar$ ), and for the critical angular momentum value 
( l$_{crit}$=55$\hbar$ )   
rather overestimate the experimental data ( see in Fig. 13 the intersection of
Bass region II and region III lines). 

At higher energies, V$_C$/E$_{c.m.}$$\le$0.3 ( region III) incomplete fusion
processes can be seen for all three systems,  becoming more important with
increasing entrance-channel mass asymmetry. In this higher energy region, a
strong fission component can be assumed for the $^{35}$Cl+$^{24}$Mg reaction
which should  be taken into account  when determining the complete fusion cross
section.  However no fission data are at present available for the two other
entrance channels, which makes interesting the question of how this process
might influence the fusion cross sections deduced for these systems. On the
other hand, although there is  some dispersion in the experimental data for the
different systems, it is clear that the complete fusion cross sections decrease
with increasing energies in region III, suggesting that a common limit is
reached for the formation of the $^{59}$Cu compound system.  Also,  this limit
appears to be independent of the specific entrance channel. In Fig. 14, the
same data are shown in a plot where the excitation energy of the compound
nucleus E$^{*}$ is shown as a function of the critical angular momentum of the
compound nucleus, as extracted from the data using the sharp cut-off
approximation. 

The saturation value found for the critical angular momentum,
J$_{cr}$$\approx$42$\pm$3.5$\hbar$, is consistent with the value at which the
fission barrier of the $^{59}$Cu compound nucleus vanishes, as calculated by
the Sierk model [48] and shown as a dashed line in the figure. Similar
behaviour has been observed for reactions leading to lighter compound nuclei
such as $^{40}$Ca [51], $^{47}$V [8], $^{56}$Ni [52] and $^{68}$Se [53,54], and
is in agreement with the systematics of Beck and Szanto de Toledo [55].

VII. SUMMARY AND CONCLUSIONS  \ \

 The properties of fusion and competing binary mechanisms have been
investigated by the  $^{35}$Cl+$^{24}$Mg reaction in the region III at an 
incident beam energy E$_{Lab}$$\simeq$282 MeV, by measuring the inclusive 
energy spectra and angular distributions of the emitted fragments which have
been identified in charge. 
The ratio of the incomplete fusion  to the complete fusion  contributions,
$\sigma_{ICF}$/$\sigma_{CF}$$\approx$0.05 to 0.10, has the same value that 
those found for the  $^{32}$S+$^{27}$Al reaction at similar bombarding energies,
 and therefore consistent with the systematics of  Morgenstern et al.[10].

In the binary-decay elemental distributions the general inelastic contribution  
evolves from quasi-elastic to full energy  damping as the charges of fragments 
are more and more remote from target and projectile Z-values.

The partially-damped exit-channels show  both  projectile-like and target-like
components, with decay-time much shorter than the rotation time 
(0.03$\le$$\tau_{1}$/T$\le$0.20). The total amount of these processes  
(sum of  the fragment contributions up to Z=16) can be evaluated to be 
$\approx$335 mb.
The fully-damped components of energy spectra can be associated with decay
times which are longer than the rotation time, therefore exhibiting  their 
clear long-lived nuclear origin.  These components, of constant
$d\sigma/d\Theta$ angular distributions, have elemental cross sections and
total kinetic energies ( corrected for secondary light-particle evaporation)
which are well reproduced by the Transition State Model of ref. [17]. 
The total cross section for this component is found to be 
$\sigma_{FF}$$\approx$168 mb. An alternative explanation for these
fully damped yields in terms of a di-nucleus orbiting mechanism cannot 
be discounted, although there is no compelling evidence for this alternative 
based on the observed systematics.
The other binary-decay mechanisms
which  compete strongly with fusion-evaporation ( $\approx$600 mb), or with
fusion followed by evaporation or fission ($\approx$768 mb), are the 
discussed damped processes ($\approx$335 mb ), the incomplete fusion process, 
( $\approx$60 mb ), and a quasi-elastic mechanism,
which  could not be  evaluated with the current  experiment.
Although with some dispersion, the  experimental data clearly indicate 
a saturation effect in region III which corresponds to a
common angular-momentum limitation reached for the formation of the $^{59}$Cu
compound system.
                                                                          
Based on the present results, it would now be interesting to revisit
the previously measured $^{32}S+^{27}Al$ and $^{19}F+^{40}Ca$ systems to more 
clearly
establish the compound nucleus origin of the fully energy damped, binary
decay yields.  This would also help to establish the significance of the
observed scatter in total fusion cross sections for the three systems at
higher energies.

\bigskip
\bigskip

     ACKNOWLEDGMENTS

     The authors wish to thank the Post-accelerated Tandem Service,
     for the kind hospitality and the technical support received 
during the
     experiment at the Saclay Laboratories.In particular they would 
like
     to mention the help of Mr. J. Girard.
     Mr. S. Leotta and Mr. S. Reito of University of Catania and
     Istituto Nazionale di Fisica Nucleare di Catania,(INFN),  
respectively,
      are also warmly thanked for their assistance during the 
experiment.
\bigskip

Footnote $^{a}$
\noindent

        See [20],  and references therein for light-particle optical
 potential
         and level density parameters


\vfill
\eject

%
%

\begin{figure}
 \caption{ Comparison of the experimental angular distribution for elastic 
 scattering to Rutherford contribution,  to optical model predictions.
Experimental values are indicated by diamond symbols.
Calculations have been performed with  the code PTOLEMY. }
\label{ Figure 1.}
\end{figure}

\begin{figure}
 \caption{ 
Energy spectra of fragments with Z = 24 - 16, measured at
$\theta_{lab}$=7$^{o}$.
To provide a more sensitive view of the results, data
size for Z=20, 19, 18 and Z=17 have been scaled by the factors  
indicated in the
figure. The arrow positions indicate the expected centroid values from eq.3.
ER-spectra (CF+ICF contributions) calculated by the LILITA  statistical-model code are 
indicated
by the full-line histograms for fragments with Z=24 to Z=18; The ICF component
is indicated by the dashed-histograms for fragments with Z=21
to Z=18. }
\label{ Figure 2.}
\end{figure}

\begin{figure}
 \caption{ 
 Energy spectra of Z=20 fragments , measured at
$\theta_{lab}$=4$^{o}$, 7$^{o}$,13$^{o}$ and 15$^{o}$, respectively.
The $\theta_{lab}$=4$^{o}$ spectrum  exhibits a large energy threshold due 
to the  identification  energy-limit of the Bragg detector used for most 
forward angles. 
Diamond symbols and histograms indicate the CF and the ICF contributions, 
respectively.
 }
\label{ Figure 3.}
\end{figure}

\begin{figure}
\caption{
Angular distributions of fragments with Z=24 to Z=16. The plus symbol are 
experimental data including all contributions. For fragments with Z=24 
to Z=22 total yields are attributed to the CF process and are compared
with the corresponding LILITA calculations (histograms). For fragments with Z=21
and Z=20  the experimental yields are attributed to the fusion processes, 
( CF+ICF ). For Z=19 and Z=18 the fusion contributions have been extracted from
total yields as indicated by diamond symbols joined by dashed lines. For 
fragments with Z=21 to Z=18 the ICF contributions are shown  by squared symbols.
}
\label{ Figure 4.}
\end{figure}

\begin{figure}
\caption{
  Comparison between the experimental complete fusion-evaporation elemental 
distributions (full circles) and the predictions of
 the LILITA and CASCADE statistical-model codes.
Empty histograms are for LILITA : small width for standard parameter
calculations [33]; large width, when ``deformation" of evaporating 
nuclei is taken
into account. Similarly for the CASCADE code : hatched histograms for 
standard
calculations [34], and  full-histograms  with ``deformation" effects, 
respectively.
}
\label{ Figure 5.}
\end{figure}

\begin{figure}
\caption{
 Comparisons between the ICF experimental data (full squares) and LILITA code 
calculations.
}
\label{ Figure 6.}
\end{figure}

\begin{figure}
\caption {
Contour plots of double differential cross sections versus TKE 
and
$\Theta_{cm}$ for fragments with Z=16 to Z=5.}
\label{ Figure 7.}
\end{figure}

\begin{figure}
\caption{
Angular distributions d$\sigma$/d$\Theta_{cm}$ for fragments with 
Z=3,4 and Z=5,
(left side); and Z=13,14,15 and Z=16, (right side),
as a function of $\Theta_{cm}$.     
For Z=13 to Z=16 and $\Theta_{cm}$$\le 
$18.5$^o$ (grazing value), data are mainly due to the positive angle, 
(near-side) contributions. 
Curves are fits to the data as described in the text.
}
\label{ Figure 8.}
\end{figure}

\begin{figure}
\caption{
The same as in Fig. 8, for Z=6 to Z=12. The fully- 
and the
partially-damped components are reported on the left and on the right side,
respectively. Energies up to 25-30 MeV have been considered 
 for fully-damped components and energies from 
25-30 MeV up to the largest energies for the partially-damped ones }
\label{ Figure 9.}
\end{figure}

\begin{figure}
\caption{ Measured total Z-distribution (full circles).
For Z=3 to Z=17, the dashed line histogram is the contribution of the fully 
damped component obtained by integration of the isotropic 
component
of d$\sigma$/d$\Theta_{cm}$ .  The full-line histogram represents the
experimental complete fusion-evaporation Z-distribution.} The excess yield 
observed around Z=16 is due to important contribution from quasi-elastic or 
partially damped collisions (see text).
\label{ Figure 10.}
\end{figure}

\begin{figure}
\caption{
Measured elemental distribution (full circles) for CF and completely damped
processes. The open squares (scaled down by a factor 0.1) are the fission
yields predicted by the FF model before evaporation. The crosses are the
fission yields (also scaled down by a factor of 0.1) after evaporation (using
the code LILITA). The calculated Z-distribution for evaporation residue (as
given by LILITA) and fission fragments (as given by the FF model after
evaporation) is shown by the full line histogram in very good agreement with
the data.} 
\label{ Figure 11.}
\end{figure}

\begin{figure}
\caption{
Average  total kinetic energies of the fragments for completely damped 
processes as a function of their charge Z. Experimental values 
(crosses with error bars are compared with the fusion-fission model 
predictions: before evaporation(open diamonds), after correction for 
evaporation (open squares).
}
\label{ Figure 12.}
\end{figure}

\begin{figure}
\caption{
Fusion excitation functions versus V$_{C}$/E$_{c.m.}$ for different reactions 
leading to the same compound nucleus $^{69}$Cu. Complete fusion 
component (full circles) and complete plus incomplete (open circles) fusion  
for the $^{19}$F+$^{40}$Ca reaction [22]. Complete fusion (open 
triangles [22], full diamonds [23], crosses [24], full square [25], open 
squares [26]) and complete plus incomplete fusion (inverted open triangles) 
for the $^{32}$S+$^{27}$Al reaction [27]. Complete fusion-evaporation 
(diamonds [20], + symbol, present work) and total ([+]symbol, present work) 
complete fusion  (fusion-evaporation and fission) for the $^{35}$Cl+$^{24}$Mg 
reaction.
The total reaction cross section, calculated with the optical model code 
PTOLEMY for the reaction $^{35}$Cl+$^{24}$Mg is indicated by the long-dashed 
line.The three intersecting straight lines represent region I, and Bass 
estimations for region II and region III respectively. }
\label{ Figure 13.}
\end{figure}

\begin{figure}
\caption{
Excitation energy E$^{*}$ versus experimental critical angular 
momenta for the
$^{59}$Cu CN. The J$_{cr}$ values were extracted from the measured
$^{32}$S+$^{27}$Al,  $^{19}$F+$^{40}$Ca and
$^{35}$Cl+$^{24}$Mg excitation functions (see text).
The symbols represent the same measurements as in Fig. 13.
The dashed line represents the disapperance of the fission barrier.
The full line is the statistical yrast line given by [ 56,49] :
E$^{*}$=E$_{c.m.}$+$\Delta$Q=[J(J+1)$\hbar$$^{2}$]/2$\Im$, with
$\Im$=(2/5)R$_{o}$A$^{5/3}_{CN}$, $\Delta$Q=10 MeV and R${_0}$=1.2 fm.
}
\label{ Figure 14.}
\end{figure}


%
\vfill
\eject
\begin{table}[p]
\caption{Binary fragment cross sections and average c.m. kinetic energies, for 
both,{\bf(a)}:  fully-damped,  and  {\bf(b)}: partially-damped,  
components. Cross section values are obtained by integration of
angular distributions reported in Fig. 8 and Fig. 9. The kinetic 
energies, TKE values, are given in MeV. The decay times, 
$\tau_{1}$ and $\tau_{2}$ are reported in units of the rotation time, 
T$\approx$1.6$\cdot$10$^{-21}$s, (see text). }
\label{ TABLE I}
\begin{tabular}{|r|r|r|r|r|r|r|r|}           
\hline
  Z     & $\sigma_{a}(mb)$  & $\sigma_{b}(mb)$  & (TKE)$_{a}$  &
 (TKE)$_{b}$  & $\tau_{1}$/T  & $\tau_{2}$/T\\
\hline
      3     & 7.2 $\pm$ 1.5  &       -       & 15.5$\pm$2  &     -      
&     -   
  &     -  \\ 
      4     & 5.9 $\pm$ 0.9  &       -       & 20.0$\pm$2  &     -       
&     -   
  &     -  \\ 
      5     & 9.7 $\pm$ 3.0  &       -       & 22.5$\pm$2  &     -       
&     -   
  &     -  \\
      6     &26.9 $\pm$ 4.4  & 10.4 $\pm$1.8 & 21.0$\pm$2  & 28$\pm$ 2   
&     -   
  &0.19$\pm$$^{0.02}_{0.02}$ \\ 
      7     &14.3 $\pm$ 2.5  &  7.0 $\pm$1.2 & 22.0$\pm$2  & 32$\pm$ 2   
&     -   
  &0.16$\pm$$^{0.02}_{0.02}$ \\ 
      8     &17.2 $\pm$ 3.2  &  9.3 $\pm$1.6 & 22.0$\pm$2  & 31$\pm$ 3   
&     -  
  &0.14$\pm$$^{0.02}_{0.02}$ \\ 
      9     &8.5  $\pm$ 1.4  &  5.2 $\pm$0.9 & 22.0$\pm$2  & 31$\pm$ 3   
&     -   
  &0.12$\pm$$^{0.02}_{0.02}$ \\ 
     10     &15.2 $\pm$ 1.8  &  8.5 $\pm$1.4 & 22.0$\pm$2  & 31$\pm$ 3   
&     -  
  &0.10$\pm$$^{0.01}_{0.02}$ \\ 
     11     &14.1 $\pm$ 2.5  &  7.6 $\pm$1.3 & 24.0$\pm$2  & 33$\pm$ 2   
&0.04$\pm$$^{0.02}_{0.01}$  
  &0.07$\pm$$^{0.01}_{0.02}$ \\ 
     12     &26.6 $\pm$ 4.8  & 10.0 $\pm$1.5 & 22.0$\pm$2  & 33$\pm$ 2   
&0.03$\pm$$^{0.02}_{0.01}$  
  &0.07$\pm$$^{0.01}_{0.03}$ \\ 
     13     &8.8  $\pm$ 1.8  & 26.6 $\pm$4.2 & 22.0$\pm$2  & 30$\pm$ 2   
&0.08$\pm$$^{0.02}_{0.02}$  
  &  -    \\ 
     14     &14.0 $\pm$ 2.8  & 61.3 $\pm$12. & 21.0$\pm$2  & 28$\pm$ 2   
&   0.08$\pm$$^{0.02}_{0.01}$  
  &   -   \\ 
     15     &8.0  $\pm$ 1.6  & 57.6 $\pm$10. & 19.0$\pm$2  & 40$\pm$ 3   
&   0.05$\pm$$^{0.02}_{0.01}$  
  &   -   \\ 
     16     &20.0 $\pm$ 4.0  &130.  $\pm$20. & 15.5$\pm$2  & 45$\pm$ 4   
&   0.04$\pm$$^{0.02}_{0.01}$  
  &   -   \\
\hline
\end{tabular}
\end{table}

\end{document}